\documentclass[fleqn, 12pt]{article}
\usepackage{multirow}
\usepackage[utf8]{inputenc}
\usepackage[T1]{fontenc}
\usepackage[english]{babel}
\usepackage{amsmath,amssymb}
\usepackage{graphicx}
\usepackage{booktabs}
\usepackage{siunitx}
\usepackage{microtype}
\usepackage{lineno}
\usepackage{newtxtext,newtxmath} 
\usepackage[numbers,sort&compress,super]{natbib}
\usepackage[paper=a4paper, margin=1.5cm]{geometry}
\usepackage{url}

\title{The Extended Ultrahigh-energy Gamma-Ray Emission in the Vicinity of PSR J2238+5903}
\author{The LHAASO Collaboration\thanks{Corresponding author: xurenfeng@ihep.ac.cn, wusha@ihep.ac.cn, lihuicai@ihep.ac.cn}\\
(The LHAASO Collaboration authors and affiliations are listed after the references.)} 

\usepackage{hyperref}

\begin{document}
\maketitle

\abstract{
We present a comprehensive analysis of the recently discovered TeV gamma-ray source, LHAASO~J2238+5900. Based on data collected from the LHAASO, our fitting results suggest that the source is significantly extended with an angular extension of $0.54^{\circ}\pm0.01^{\circ}$ and is spatially coincident with the pulsar PSR J2238+5903. Its spectrum is characterized by a power-law with a cutoff at $41.0\pm3.5$\,TeV. Additionally, the source exhibits a significant signal of 7.9$\sigma$ above 100\,TeV, implying that it is a PeVatron candidate. While the gamma-ray emission is consistent with a pulsar wind nebula (PWN) scenario, the relatively large extension size also allows for a halo interpretation, potentially caused by electron-positron pairs escaping from the PWN.}


\section{Introduction} \label{sec:intro}
As observational sensitivity improves, an increasing number of extended gamma-ray sources have been detected across the sky \citep{2023arXiv230712546B, 2020ApJ90576A, 2018A&A612A2H, abdalla2018hess,2024ApJS27125C}. These large sources exhibit significant interactions with their surrounding environments. Currently, the majority of large extended ultrahigh-energy (UHE, photon energy $E\geq 100\,\mathrm{TeV}$) galactic sources are associated with pulsar wind nebulae (PWNe) and TeV Halos  \citep{2025J0248,LHAASOCTA1,LHAASOJ1740,LHAASOJ1849}. Young PWNe in their early expansion phase exhibit strong broadband emissions across multiple wavelengths, though confined within relatively compact physical regions. As these nebulae evolve, the reverse shock from the supernova remnant (SNR) moves inward toward the pulsar. Due to the asymmetric nature of the initial explosion and inhomogeneities in the surrounding medium, this process often distorts the morphology of the PWN, resulting in irregular and asymmetric structures—such as those observed in Vela-X \citep{aharonian2006first,de2012probing}. These young PWNe typically display energy-dependent morphological features. For instance, in sources like HESS J1825-137 and HESS J1303-631 \citep{2006A&A460365A, 2019A&A621A116H, 2012A&A548A46H}, the emission region appears to shift toward the pulsar with increasing photon energy, reflecting the historical imprint of lower-energy electron-positron pairs left behind by earlier outflows. In addition, the spatial extent of the emission region tends to decrease significantly at higher energies, a behavior reminiscent of X-ray observations, suggesting that advection dominates the particle transport process in such systems. In contrast, TeV halos such as Geminga, Monogem, and LHAASO J0622+3754 \citep{2017Sci358911A,2021PhRvL126x1103A} also exhibit large angular extensions. However, the diffuse morphology of these halos, coupled with the extended particle distributions within the surrounding interstellar medium (ISM), implies that they are powered by middle-aged pulsars rather than young ones. Despite these observations, the number of well-studied examples of such large and extended gamma-ray sources remains limited. This scarcity hinders a comprehensive understanding of the underlying mechanisms governing particle acceleration, transport dynamics, and evolutionary pathways in PWNe and related systems.

Since becoming operational in 2021, the Large High Altitude Air Shower Observatory (LHAASO), one of the most sensitive TeV instruments, has uncovered numerous new faint extended sources. The newly discovered extended gamma-ray source LHAASO~J2238+5900, reported in the First LHAASO Catalog \citep{2024ApJS27125C}, is suggested linked to PSR~J2238+5903 with a low chance probability of about 0.03\%. PSR J2238+5903 is considered the most likely candidate, given its high spin-down luminosity of approximately $8.9 \times 10^{35}\ \mathrm{erg\ s^{-1}}$ and a characteristic age of 26.6\,kyr. However, it exhibits low radio flux ($\rm{S_{1400}<0.011\,mJy}$) \citep{ray2011precise}. In the X-ray band, an observation of the pulsar was conducted by the XMM-Newton telescope (Obs.ID=13286), but due to a short exposure time of just 9.84 ks, no extended nebula was detected. In the GeV regime, the pulsar exhibits clear pulsed emission detected by Fermi-LAT, which is cataloged as 4FGL J2238.5+5903 \citep{2009Sci325840A,2020ApJS24733A}. In the TeV regime, several TeV observatories have followed up on Fermi results. The MAGIC collaboration conducted a 44-hour observation targeting PSR J2238+5903 to search for a potential PWN, but no significant excess was observed. An upper limit on the differential flux was obtained at the 95\% confidence level, approximately $\sim 10^{-13}\,\rm{TeV^{-1}\,cm^{-2}\,s^{-1}}$ at a few TeV \citep{barral2018extreme}. The Tibet $AS\gamma$ experiment reported faint indications of a possible TeV emission, with a 2.5$\sigma$ hint near the position of the pulsar \citep{amenomori2009observation}. The MILAGRO collaboration observed a higher significance of 4.7$\sigma$ \citep{2014APh5716A}, which still remains insufficient to support a detailed analysis. Additionally, the extended gamma-ray region contains another pulsar, PSR J2240+5832, which exhibits a high spin-down luminosity of $2.2 \times 10^{35}\ \mathrm{erg\ s^{-1}}$ and a characteristic age of 144\,kyr. However, given its relatively large offset of approximately $0.6^\circ$ from the emission center, it is not considered a prime candidate.

In this study, we present a detailed analysis of the LHAASO~J2238+5900 using LHAASO data. The paper is structured as follows: Section \ref{sec:ins_data} describes the instrument and data analysis methods utilized in the study. Section \ref{sec:result} presents the analysis results of LHAASO J2238+5900. In Section \ref{sec:discussion}, the emission origin and broadband spectral analysis of the source are discussed. Finally, Section \ref{sec:conclusion} summarizes the conclusions drawn from the study.

\section{Instrument and Data analysis method}\label{sec:ins_data}
As a ground-based cosmic ray observatory, LHAASO is located on Mount Haizi in Daocheng, Sichuan Province, China, at an altitude of 4,410\,m. It is a hybrid observatory consisting of a 1.3\,$\rm{km}^2$ array (KM2A), a Water Cherenkov Detector Array (WCDA), and wide field-of-view air Cherenkov telescopes (WFCTA). Each sub-array and telescopes can work independently or in combination to observe air showers, such as PeV events from the Crab Nebula \citep{2021Sci373425L}. The sub-arrays are designed to cover distinct energy bands, enabling gamma-ray detection from hundreds of GeV to beyond 1\,PeV. More information of the detectors can refer to \cite{he2018design}. After the secondary particles trigger the detectors, we can reconstruct the arrival direction, the core position and the energy of showers. By using the parameter $compactness$, the rejection power of WCDA can reach up to 99.8\% for large showers with 700 hits. With the help of the muon detectors (MDs), KM2A achieves good gamma-hadron discrimination, with a hadron survival rate suppressed to 0.025\% at 100\,TeV. The celestial region from $0^\circ$ to $360^\circ$ in right ascension and $-20^\circ$ to $80^\circ$ in declination is divided into spatial bins of size $0.1^\circ \times 0.1^\circ$, which are filled with the number of the detected events. The cosmic ray background events in the bin is estimated using the direct integration method. More details for the reconstruction and performance of WCDA and KM2A can refer to \cite{2021ChPhC45b5002A,2021ChPhC45h5002A}. 

Our analysis utilizes WCDA data from March 2021 to July 2025 across five bins of the number of triggered units ($N_{\text{hit}}$:100-200, 200-300, 300-500, 500-800 and $\geq$800),  together with KM2A data from December 2019 to July 2025, which cover 25\,TeV to 1\,PeV and are logarithmically binned at five bins per decade. A three-dimensional likelihood analysis is performed to model the spatial morphology and the spectral energy distribution (SED) simultaneously. The statistical significance of the source is assessed using the test statistic (TS), defined as:
$\mathrm{TS} = 2 \ln \left( \frac{\mathcal{L}_{s+b}}{\mathcal{L}_{b}} \right)$,
where $\mathcal{L}_{s+b}$ and $\mathcal{L}_{b}$ denote the maximum likelihood values under the signal-plus-background and background-only (null) hypotheses, respectively. The distribution of TS is asymptotically $\chi^2$, with the number of degrees of freedom given by the difference in the number of free parameters between the two hypotheses. In the case of a point source with a fixed position and spectral shape, where the normalization is the sole free parameter, the pretrial significance is approximately $\sqrt{\mathrm{TS}}$. In order to eliminate the influence of nearby sources, a large region of interest (ROI) is selected as a rectangular of $20^{\circ}\times10^{\circ}$, centered on the SNR G106.3+2.7  to simultaneously cover both sources. We iteratively add source with a power-law spectral shape and a 2D Gaussian spatial model to the region until the log-likelihood value shows no significant improvement, corresponding to a significance threshold of $\Delta \rm{TS} > 25$. After fixing the number of sources, we compare the 2D Gaussian and diffusion models to determine the optimal spatial model for the target source. The diffusion template is taken from  \cite{2021PhRvL126x1103A} and is represented by the equation:
\begin{equation}
    f(\theta) \propto \frac{1}{\theta_{d}(\theta + 0.085\theta_{d})}\rm{exp}[-1.54(\theta/\theta_{d})^{1.52}],
\end{equation} 
where $\theta_d = (180^\circ/\pi) \cdot 2\sqrt{D(E_{e})t_{E}}/d$ represents the diffuse size, $D(E_e)$ is the diffusion coefficient at energy $E_e$, $t_E$ take the value of $\min (t_{\rm cool}, t_{\rm age})$ and $d$ is the distance of the pulsar. In addition, a forward-folding method is used, in which simulated excess events are compared with the detected results to derive the differential energy spectrum\citep{cao2024lhaaso}.

\begin{figure}[!ht]
    \centering
    \includegraphics[width=0.96\textwidth]{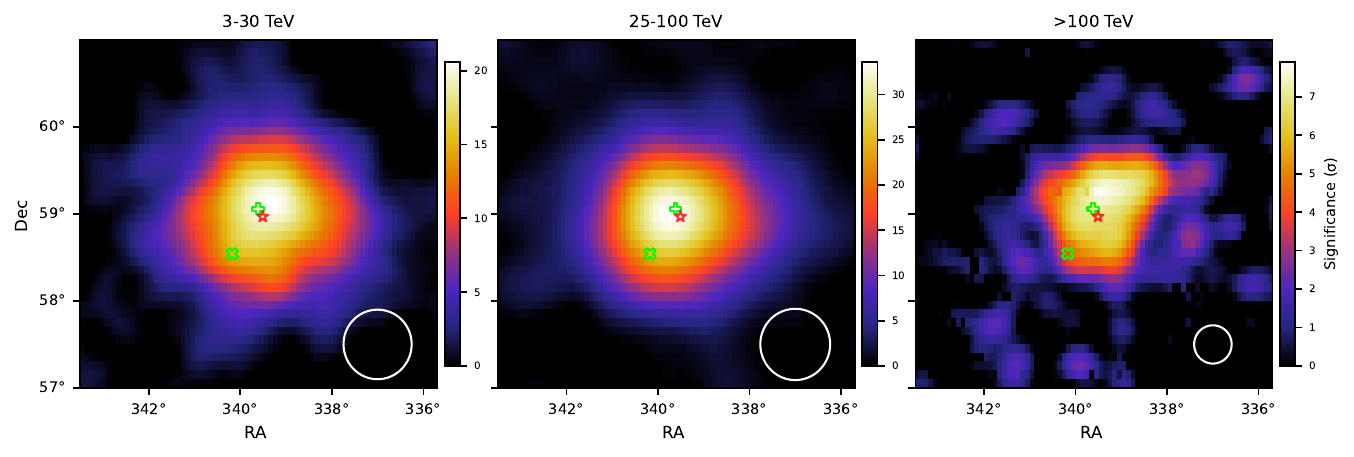}
    \caption{Significance maps of the LHAASO J2238+5900 region divided into three energy bins. The lime plus, the lime cross, and the red star represent the position of PSR~J2238+5903, PSR~J2240+5832, and LHAASO J2238+5900 (this work), respectively. The white circle at the bottom-right corner shows the size of the point-spread function (68\% containment). }
    \label{fig:tsmap}
\end{figure}

\section{Results}\label{sec:result}
Between the two tested models, the 2D Gaussian template is favored, yielding a TS value across the full ROI that is larger by approximately 42 compared to the diffusion model with $\theta_d=1.49^\circ\pm0.06^\circ$. Under the Gaussian model, the best-fit extension is $\sigma=0.54^{\circ}\pm0.01^\circ$ and the position is (RA=$339.48^{\circ}\pm0.04^{\circ}$, Decl=$58.97^{\circ}\pm0.02^{\circ}$). Evidently, LHAASO~J2238+5900 is in close proximity of PSR~J2238+5903, with an angular separation of approximately $0.14^{\circ}$, which is considerably smaller than the radius of our instrument’s PSF. We rebinned the original data into three energy intervals: 3--30\,TeV, 25--100\,TeV, and $>100$\,TeV. Figure \ref{fig:tsmap} shows the significance maps of LHAASO~J2238+5900 in the three energy intervals, subtracting emission from the diffuse backgrounds and other sources included in the best-fit model. The LHAASO~J2238+5900 is detected with a significance of $7.9\sigma$ above 100\,TeV, suggesting that it is a PeVatron candidate.

\begin{figure}[h]
    \centering   
    \includegraphics[width=0.96\textwidth]{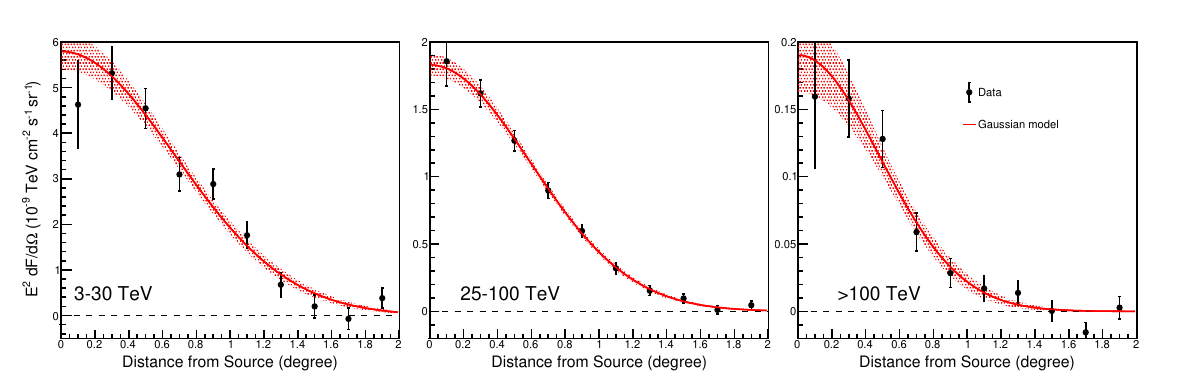}
    \caption{One-dimensional distribution of the gamma-ray emission of LHAASO J2238+5900. The solid line represents the best-fitting Gaussian model, and the shaded band is the $\pm 1\sigma$ statistical uncertainty.}
    \label{fig:Profile}
\end{figure}

\begin{figure}[h]
    \centering
    \includegraphics[width=0.5\textwidth]{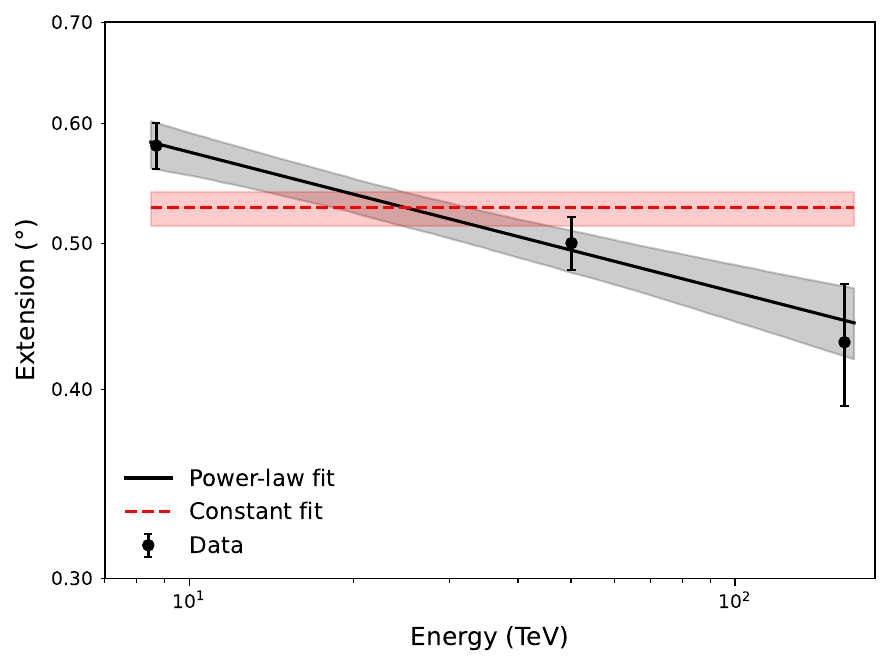}
    \caption{The measured extension ($\sigma_{\text{ext}}$) for LHAASO J2238+5900 at three energy intervals. The solid black line indicates the power-law fit, and the red dashed line indicates the constant fit result. }
    \label{fig:size}
\end{figure}

To explore the position and morphology in more detail, we performed the 3D likelihood analysis in each of the three energy intervals. The morphological and spectral parameters of other sources were held fixed, while the extensions of LHAASO~J2238+5900 and the normalizations of all sources in the ROI were free parameters. The distributions of surface brightness were profiled as a function of angular distance from the best-fit gamma-ray emission centroid, as shown in Figure \ref{fig:Profile}. Within about $1^\circ$ from the centroid, the excess events are primarily attributed to the LHAASO~J2238+5900. In our study, the gamma-ray emission centroid in the three energy intervals is measured to be offset from PSR~J2238+5903 by $0.14^{\circ}\pm0.03^{\circ}$, $0.11^{\circ}\pm0.02^{\circ}$ and $0.17^{\circ}\pm0.06^{\circ}$, respectively. Considering the statistical uncertainties, the position of the gamma-ray emission does not exhibit a significant variation with energy. In addition, the source size (the Gaussian width $\sigma_{\text{ext}}$) is measured to be $0.58^{\circ}\pm0.02^{\circ}$, $0.50^{\circ}\pm0.02^{\circ}$ and $0.43^{\circ}\pm0.04^{\circ}$, respectively. The energy-dependent extent, displayed in Figure \ref{fig:size}, is well described by a power law of the form $\sigma_{\text{ext}}=(0.71\pm0.06)\times E^{(-0.09\pm0.03)}$ ($\chi^2/ndf=0.2/1$). However, this represents only a $2.7\sigma$ improvement over a constant fit ($\chi^2/ndf=7.4/2$), suggesting that the data do not strongly favor an energy-dependent model.

\begin{figure}[h]
    \centering
    \includegraphics[width=0.5\textwidth]{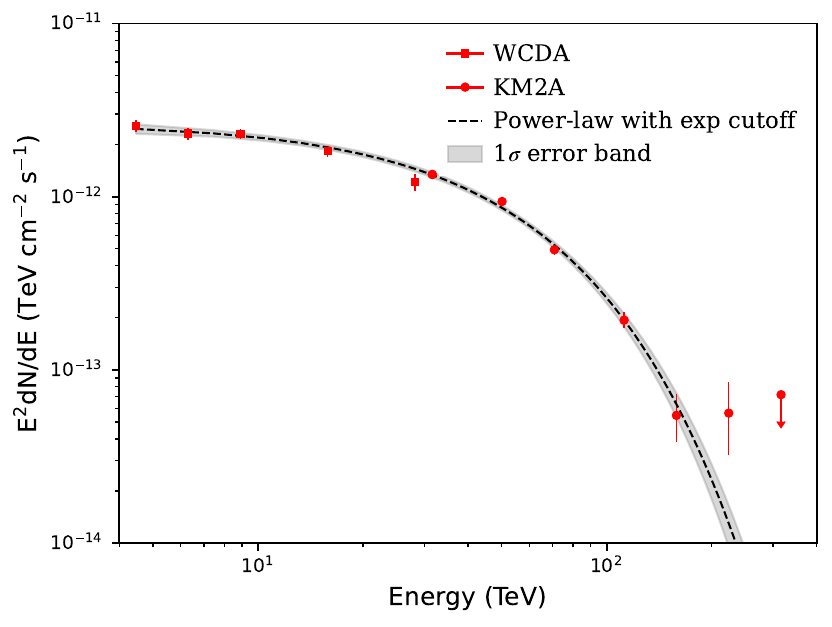}
    \caption{The SED of LHAASO~J2238+5900 and the corresponding fit result.}
    \label{fig:sed}   
\end{figure}

Using the Gaussian template, the energy spectrum in the source region is derived, with the results shown in Figure \ref{fig:sed}. The spectrum exhibits a rapid cut-off at high energies, formalized as $N_0(E/E_0)^{-\Gamma}\exp(-{E/E_{c}})$. A comparison of the fitting results demonstrates a clear preference for the energy-cutoff spectral model ($\chi^2/ndf=10.7/8$) over the simple power law ($\chi^2/ndf=343.4/9$). The best-fit parameters are as follows: $N_0 = (3.21\pm 0.28)\times 10^{-15}\,\rm{TeV^{-1}\,cm^{-2}\,s^{-1}}$ at 30\,TeV, $\Gamma=1.97\pm0.06$, and $E_{\text{c}}=41.0\pm3.5\,\text{TeV}$. The integrated flux above 1\,TeV is $2.4\times10^{-12}\,\rm{cm^{-2}\,s^{-1}}$, approximately equivalent to $12\%$ of the Crab flux. As mentioned in \cite{2024ApJS27125C}, the primary systematic error in the spectrum arises from the atmospheric model used in the Monte Carlo simulations, resulting in an influence of about $7\%$ on the flux and 0.02 on the spectral index for the KM2A data. The WCDA data exhibit a flux uncertainty of $8\%$, while the uncertainty of the spectral index varies with the shape of the energy spectrum.

\section{DISCUSSION}\label{sec:discussion}
The most likely counterpart for the gamma-ray emission from LHAASO J2238+5900 is PSR~J2238+5903. Distance estimates for this pulsar vary depending on the method used. The Third Fermi-LAT Catalog of Gamma-ray Pulsars (3PC) sets a distance limit of less than 11.3 kpc using the ``DMM'' method, which estimates the distance to a pulsar based on its dispersion measure \citep{2023ApJ958191S}. In contrast, an alternative approach, based on a phenomenological gamma-ray luminosity law, suggests a much smaller distance of 2.11 kpc \citep{brownsberger2014survey}. Recently, FAST detected radio pulsations from PSR~J2238+5903, with a distance of approximately 7 kpc estimated by the YMW16 Galactic electron density model \citep{2026Fast}, and we adopt this distance in this work. Currently, the pulsar lacks deep X-ray observations. Additionally, the Fermi-LAT observations indicate weak off-pulse emissions in this region. We derived the upper limits with the same spatial template as LHAASO J2238+5900 (see Appendix A for details). To explore the gas distribution around the region, we applied the CO data from the Milky Way Imaging Scroll Painting (MWISP) survey, a large-scale CO survey led by the Purple Mountain Observatory in China \citep{Su_2019}. Several small clouds are observed around the source; however, they are offset from the brightest region of LHAASO J2238+5900 and fail to encompass the extended emission area (see Appendix B). Additionally, the precise distances to these clouds remain unknown, making it unclear whether they are physically associated with the LHAASO source. We also searched the catalog of \cite{2017ApJ83457M} for cloud information along the line of sight to LHAASO J2238+5900, extending over $0.5^{\circ}$. Four clouds were identified at distances of 1.87, 4.97, 6.21 and 6.97 kpc which have relatively low densities, peaking at only around 12 $\rm{cm^{-3}}$. The fourth cloud, [MML2017]4246, exhibits a distance consistent with that of PSR J2238+5903 and an average angular radius of $0.36^{\circ}$. However, it shows an angular offset of $ \sim 0.5^{\circ}$ from the gamma-ray emission center (Figure \ref{fig:co_tsmap}).

\begin{figure}[h]
    \centering
    \includegraphics[width=0.5\textwidth]{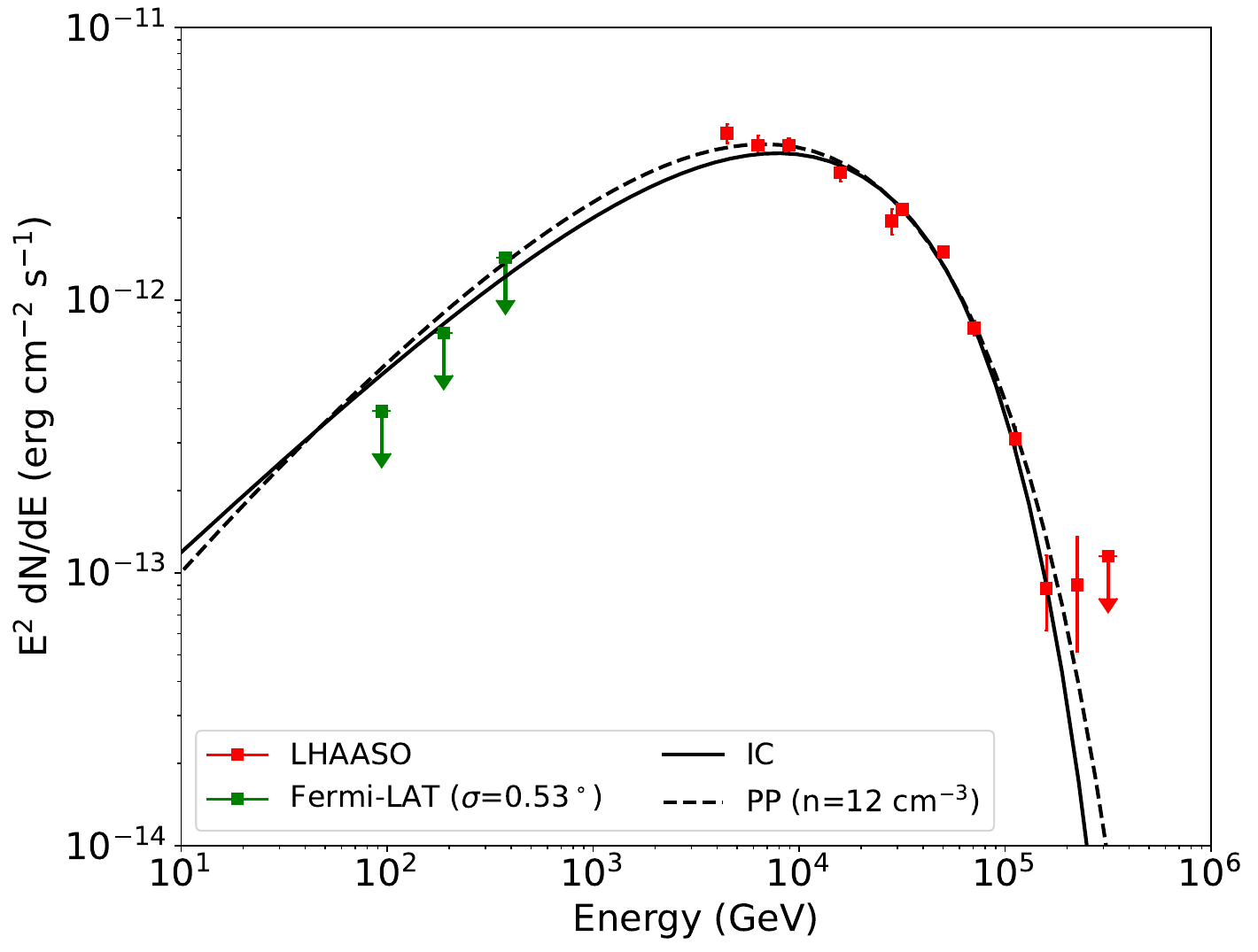}
    \caption{Multiwavelength SED of LHAASO J2238+5900 with hadronic (dashed line) and leptonic (solid line) models. The red squares represent the spectral data points from LHAASO, while the green squares indicate the upper-limit points derived from Fermi-LAT, as outlined in APPENDIX \ref{sec:fermi_result}. }
    \label{fig:mult_sed}
\end{figure}

We consider both leptonic and hadronic origins and fit the Fermi-LAT and LHAASO data with PYTHON package Naima \citep{2015ICRC34922Z,2014ApJ783100K,2010PhRvD82d3002A}. Given the presence of a cutoff in the gamma-ray spectrum, the electron and proton spectra are modeled as a power law with an exponential cutoff. In leptonic model, the gamma-ray photons are produced by the inverse Compton (IC) scattering process of electrons. We take into account three seed photon fields: the Cosmic Microwave Background (CMB), far-infrared dust emission (FIR), and near-infrared stellar emission(NIR). The energy densities of these fields are $U=0.26, 0.32, 0.31\,\rm{eV\,cm^{-3}}$, respectively. The corresponding temperatures are $T=2.7, 29.0, 5360\,\rm{K}$, which are derived from the Interstellar Radiation Field (ISRF) model proposed by \cite{2022MNRAS5092339N}. As shown in Figure \ref{fig:mult_sed}, the electrons, characterized by a spectral index of 1.55 and an exponential cutoff at 45\,TeV, can well explain the data. The total energy required for this population is $2.6\times10^{47}$ erg. In the hadronic model, we calculate the proton-proton process assuming a maximum density of 12 $\rm{cm^{-3}}$. Although a proton spectrum with an index of 1.2 and a cutoff energy of 115\,TeV can reproduce the Fermi-LAT and LHAASO data, it is substantially harder than the typical spectra of protons accelerated in SNR. This suggests that other spectral shapes for protons, or multi-zone models, may be required. Given the unresolved nature of proton acceleration efficiency in pulsars, our study concentrates primarily on the leptonic origin.

\subsection{New PWN?}
Based on the distance of 7\,kpc, the corresponding 1–30\,TeV gamma-ray luminosity is estimated to be $7.2\times 10^{34}\,\rm{erg\,s^{-1}}$, implying a gamma-ray conversion efficiency of 8\%. This value is slightly higher than the typical range for very high-energy (VHE) PWNe, where efficiencies are generally between 0\% and 7\% \citep{2009ApJ69412M}. It is plausible that the pulsar powers a PWN responsible for photon generation. Generally, PWNe exhibit notable energy evolution characteristics, as demonstrated by sources such as HESS~J1825-137 and HESS~J1303-631 \citep{2006A&A460365A, 2019A&A621A116H, 2012A&A548A46H}, where both the spatial extent of the emission significantly shrinks and the emission centroid moves progressively closer to the pulsar as the energy increases. The pulsar PSR J2238+5903 and PSR B1823-13 have comparable ages. However, the LHAASO J2238+5900's centroid does not vary significantly with energy, and its extension shows a decreasing trend that is only marginally significant. 

The extension evolution indicates that particles around LHAASO J2238+5900 may be attributable to the diffuse transport mechanism. In a diffusion-dominated scenario, the diffusion coefficient is energy-dependent, given by $D(E)=D_0(E_e/E_0)^{\delta}$, leading to the expression for the extent size as $R=\sqrt{2D\tau}$, with $\tau$ taken to be the minimum of the accelerator's age and the cooling timescale of the radiating particles. For synchrotron and IC scattering, the cooling time is assumed to scale as $\tau\propto1/E_e$. In the Thomson regime, where $E_{\gamma}\propto E_{e}^2$, we have $R\propto E_{\gamma}^{(\delta-1)/4}$ (in the Klein-Nishina regime, $E_{\gamma}\propto E_{e}$, $R\propto E_{\gamma}^{(\delta-1)/2}$). Typically, the parameter $\delta$ ranges from 0 to 1, representing energy-independent diffusion and Bohm diffusion, respectively. From the fit of the energy-dependent extent, we derive $\delta \approx 0.6$ for Thomson regime (and $\delta\approx0.8$ for the KN regime). In our scenario, with a precise pulsar age and a magnetic field of 3 $\mu$G, electrons with an energy of roughly 30 TeV would have a cooling time comparable to the pulsar's age. This implies that for lower energies, $\tau$ remains constant, corresponding to $R\propto E_{\gamma}^{\delta/4}$. Since our first energy interval is below 30 TeV, the maximum extension could be larger around this energy, resulting in a steeper fit line and a slightly smaller $\delta$ value.

\subsection{Young Halo?}
Adopting a distance of 7 kpc to the PSR J2238+5903, the physical extent enclosing 90\% of the source's emission is about 140 pc. Given its large physical scale, LHAASO J2238+5900 may be classified as a halo. Nonetheless, no significant deviations from the 2D Gaussian model are observed when the morphology is fitted using a diffusion template. This does not necessarily argue against the halo interpretation, as the morphology of pulsar halos can be influenced by various effects, such as the particle transport mechanism (e.g., isotropic diffusion, ballistic transport, and anisotropic diffusion), non-steady injection, projection effects, and contamination from unresolved sources. \citep{2019PhRvL123v1103L, 2024ApJ9699W, 2022FrASS922100F} According to the relation between electrons and inverse Compton scattering photons \citep{2017Sci358911A}, 20\,TeV photons correspond to electrons with an energy of approximately 100\,TeV. The derived diffusion coefficient for 100 TeV electrons, estimated at $9.4\times10^{28}(d/7\ \text{kpc})^2\ \text{cm}^2\ \text{s}^{-1}$, is approximately one order of magnitude higher than those inferred for other well-known TeV Halos, including Geminga, PSR B0656+14 \citep{2017Sci358911A}, and LHAASO J0621+3755 \citep{2021PhRvL126x1103A}. Since the inferred diffusion coefficient scales quadratically with the distance, this discrepancy may be attributed to the uncertainty in the distance measurement. Alternatively, it could also be explained by the young nature of this halo, where the diffusion-suppression mechanism has not yet fully developed.

An interesting aspect of this system is the relative youth of PSR J2238+5903 compared to other pulsars that power TeV Halos. The TeV Halos observed thus far are typically associated with middle-aged pulsars, which have likely exited their SNRs and formed distinctive bow-shock PWNe. These electrons escape from the nebula and begin diffusing throughout the surrounding ISM. Once the pulsar reaches a distance of roughly 0.68 times the SNR radius, its  velocity surpasses the local sound speed. Pulsars such as W44 and the Vela pulsar, which are still embedded within their respective SNRs, are believed to have already developed BSPWN structures, as proposed by \cite{2011ApJ740L26C}. In these cases, the high velocity of the pulsars or the distortion from the reverse shock can cause electrons to disperse or escape \citep{2019ApJ87754B, 2019ApJ881148B, 2025ApJ98098S}. The search findings from the HAWC Collaboration revealed some extended TeV sources surrounding young systems \citep{2017PhRvD96j3016L}. For instance, PSR J0359+5414, a relatively young and radio-quiet pulsar, is believed to power a large extended TeV Halo \citep{2023ApJ944L29A}. Lastly, although no emission from the associated SNR has been detected in other wavelengths, the possibility of SNR contamination affecting the LHAASO gamma-ray source cannot be ruled out.

\section{Conclusions}\label{sec:conclusion}
In this work, we present the precise measurements of the newly-discovered UHE gamma-ray source LHAASO J2238+5900. A Gaussian fit indicates a spatial extension of $0.54^{\circ}\pm0.01^{\circ}$. It is located at an angular distance of $0.14^{\circ}$ from PSR J2238+5903. Given its high spin-down luminosity, the pulsar is most likely associated with LHAASO J2238+5900. Multiwavelength observations yield no significant counterpart emissions, and there is a lack of compelling evidence for dense molecular or atomic material enveloping the gamma-ray source. A leptonic origin can naturally account for the observed photons. The high-energy electrons responsible for the inverse Compton scattering could be supplied by either a PWN or a TeV halo. To fully unravel the intricacies of the transport mechanism, it will be crucial to obtain data with enhanced spatial resolution, such as that delivered by Cherenkov telescopes.

\noindent {\bf Acknowledgements}
We would like to thank the staff members of National Facility for LHAASO (CSTR: \url{https://cstr.cn/31117.02.LHAASO}) who work at the LHAASO site above 4400 meter above the sea level year round to maintain the detector and keep the water recycling system, electricity power supply and other components of the experiment operating smoothly. We are grateful to Chengdu Management Committee of Tianfu New Area for the constant financial support for research with LHAASO data. We appreciate the computing and data service support provided by the National High Energy Physics Data Center for the data analysis in this paper. This research work is supported by the following grants: The National Natural Science Foundation of China No.12393851, No.12393852, No.12393853, No.12393854, No.12205314, No.12261160362, No.12305120, No.12375107, No.12522510, No.12575120, National Key Research and Development Program of China 2025YFE0202600, Department of Science and Technology of Sichuan Province No.24NSFJQ0060, No.2024ZYD0111, Project for Young Scientists in Basic Research of Chinese Academy of Sciences No.YSBR-061 and in Thailand from the NSRF via the Research and Innovation Acceleration Agency for Competitiveness and Area Development (RCAD) (Program Management Unit for Technology and Innovation for Future Industries (PMU-B) : Brainpower for Future Industries) [grant number B39G690003].

This research made use of the data from the Milky Way Imaging Scroll Painting (MWISP) project, which is a multi-line survey in $^{12}$CO/$^{13}$CO/C$^{18}$O along the northern galactic plane with PMO-13.7 m telescope. We are grateful to all the members of the MWISP working group, particularly the staff members at PMO-13.7 m telescope, for their long-term support. MWISP was sponsored by National Key R\&D Program of China with grants 2023YFA1608000 \& 2017YFA0402701 and by CAS Key Research Program of Frontier Sciences with grant QYZDJ-SSW-SLH047.

\noindent {\bf Author Contributions}
R.F. Xu and S. Wu prepared the drafting of the manuscript and performed the data analysis for KM2A. R.F. Xu conducted the Fermi-LAT data analysis and MWISP analysis. Huicai Li provide the WCDA data result. We thank S.Q. Xi and S.Z. Chen for providing the analysis tools and valuable suggestions on the analysis. We would like to express our gratitude to R. Wang for her cross check for WCDA result, H.M. Zhang for assistance with the analysis and Q.Y. Cheng for carefully reading the manuscript and improving the English. The whole LHAASO collaboration contributed to the publication, with involvement at various stages ranging from the design, construction, and operation of the instrument, to the development and maintenance of all software for data calibration, data reconstruction, and data analysis. 

\appendix
\section{Fermi-LAT Analysis} \label{sec:fermi_result}
Photon data files with timing analysis results for the PSR J2238+5903 are provided in 3PC, covering period from August 4, 2008, to December 22, 2019 \citep{2023ApJ958191S}. The data covers the energy range from 20 MeV to 1 TeV, centered on the pulsar. Events were filtered out based on a zenith angle cut ($> 105^{\circ}$) and poor data quality flags. In this analysis, we use Python package $FermiPy$ to analysis the same data. The ROI was selected as $15^\circ\times15^{\circ}$ region centered at the position of 4FGL~J2238.5+5903. The source models are constructed based on the latest Fermi Large Area Telescope Fourth Source Catalog \citep{2023arXiv230712546B}. Additionally, the Galactic and extra-galactic diffuse emissions were obtained from gll\_iem\_v07.fits and iso\_P8R3\_SOURCE\_V3\_v1.txt, respectively. Since the PSR~J2238+5903 is a bright gamma-ray pulsar, we adopt an off-pulse analysis, employing the same phase (0.375-0.5 and 0.625-1.0) defined in \cite{2024ApJ968117Z}. Therefore, we derived the upper limit flux with the same spatial template as LHAASO J2238+5900 ($\sigma=0.5^\circ$). The obtained results are shown in Figure \ref{fig:mult_sed}.

\section{MWISP Analysis}
\begin{figure}[h]
    \centering
    \includegraphics[width=0.8\textwidth]{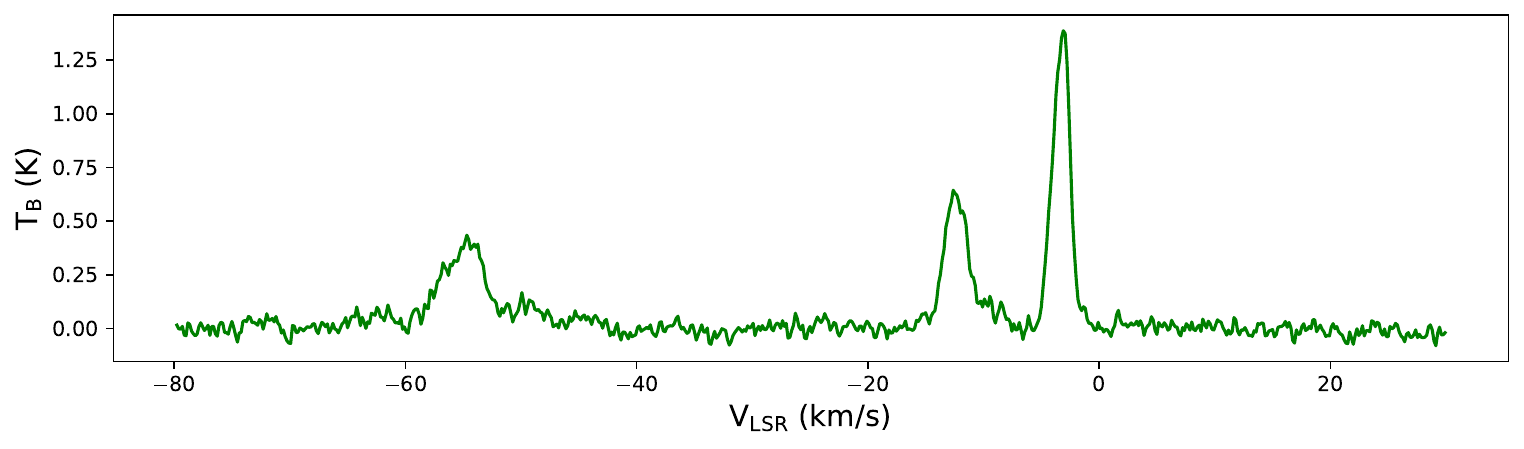}
    \caption{${}^{12}\rm{CO}$ spectrum over the selected region.}
    \label{fig:vlsr}
\end{figure}

\begin{figure}[h]
    \centering
    \includegraphics[width=0.98\textwidth]{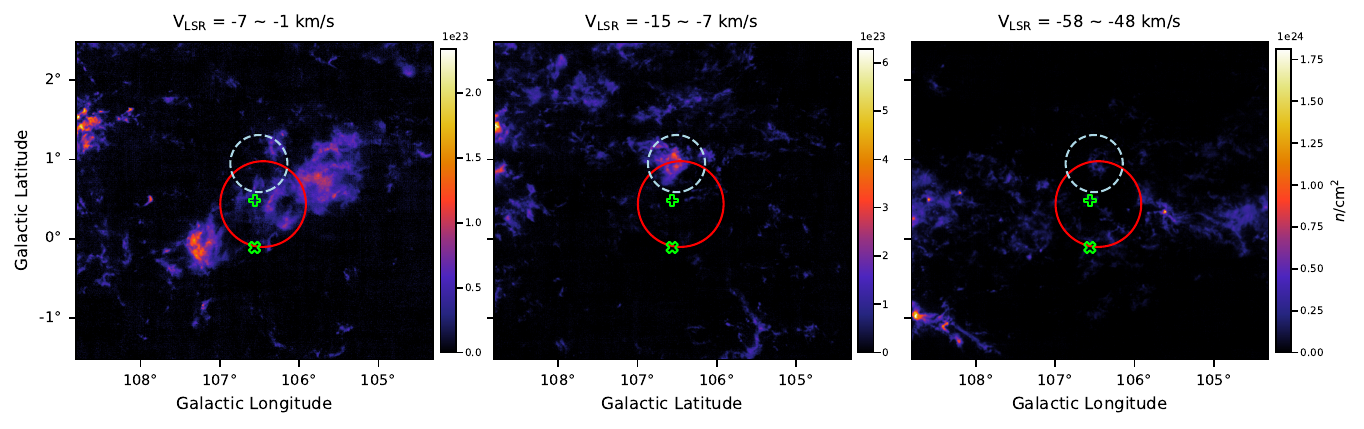}
    \caption{Column density of molecular gas toward LHAASO J2238+5900. The extended region of LHAASO is marked by solid circle. The lime plus and cross represent the position of PSR~J2238+5903 and PSR~J2240+5832, respectively. The light blue dashed circle represents the molecular cloud [MML2017]4246.}
    \label{fig:co_tsmap}
\end{figure}

The MWISP observation survey covers most of the Galactic plane at a low galactic latitude ($|b|<5^{\circ}$). It includes data from three CO isotopologue lines: ${}^{12}\rm{CO(J=1\rightarrow 0})$, ${}^{13}\rm{CO(J=1\rightarrow 0})$, and $\rm{C{}^{18}O(J=1\rightarrow 0})$. The ${}^{12}\rm{CO}$ line has a noise level of approximately 0.5\,K at a velocity resolution of 0.16$\,\rm{km\,s^{-1}}$, while the ${}^{13}\rm{CO}$ and $\rm{C^{18}O}$ lines have a noise level of approximately 0.25\,K at 0.17$\,\rm{km\,s^{-1}}$. In this work, we select a region of $4.5^{\circ} \times 4.0^{\circ}$ center on the PSR~J2238+5903. Along the light of sight, three distinct velocity intervals $(-7\sim-1),( -15\sim-7)$ and $(-58\sim -48)\,\rm{\,km\,s^{-1}}$ are above the background level (Figure \ref{fig:vlsr}). The cloud\rq s column density are shown in Figure \ref{fig:co_tsmap}.

\bibliographystyle{unsrt}
\bibliography{reference}

\begin{thebibliography}{10}

\bibitem{2023arXiv230712546B}
J.~{Ballet}, P.~{Bruel}, T.~H. {Burnett}, et~al.
\newblock {Fermi Large Area Telescope Fourth Source Catalog Data Release 4
  (4FGL-DR4)}.
\newblock {\em arXiv e-prints}, July 2023.

\bibitem{2020ApJ90576A}
A.~{Albert}, R.~{Alfaro}, C.~{Alvarez}, et~al.
\newblock {3HWC: The Third HAWC Catalog of Very-high-energy Gamma-Ray Sources}.
\newblock {\em ApJ}, 905(1):76, December 2020.

\bibitem{2018A&A612A2H}
H.~{Abdalla}, A.~{Abramowski}, F.~{Aharonian}, et~al.
\newblock {The population of TeV pulsar wind nebulae in the H.E.S.S. Galactic
  Plane Survey}.
\newblock {\em Astronomy \& Astrophysics}, 612:A2, April 2018.

\bibitem{abdalla2018hess}
H.~{Abdalla}, A.~{Abramowski}, F.~{Aharonian}, et~al.
\newblock {The H.E.S.S. Galactic plane survey}.
\newblock {\em Astronomy \& Astrophysics}, 612:A1, April 2018.

\bibitem{2024ApJS27125C}
Zhen {Cao}, F.~{Aharonian}, Q.~{An}, et~al.
\newblock {The First LHAASO Catalog of Gamma-Ray Sources}.
\newblock {\em ApJS}, 271(1):25, March 2024.

\bibitem{2025J0248}
Zhen {Cao}, F.~{Aharonian}, {Axikegu}, et~al.
\newblock {LHAASO detection of very-high-energy {\ensuremath{\gamma}}-ray
  emission surrounding PSR J0248+6021}.
\newblock {\em Science China Physics, Mechanics, and Astronomy}, 68(7):279504,
  July 2025.

\bibitem{LHAASOCTA1}
Zhen {Cao}, F.~{Aharonian}, {Axikegu}, et~al.
\newblock {Deep view of composite SNR CTA1 with LHAASO in
  {\ensuremath{\gamma}}-rays up to 300 TeV}.
\newblock {\em Science China Physics, Mechanics, and Astronomy}, 68(7):279503,
  July 2025.

\bibitem{LHAASOJ1740}
Zhen {Cao}, F.~{Aharonian}, Y.~X. {Bai}, et~al.
\newblock {Ultra-high-energy $\gamma$-ray emission associated with the tail of
  a bow-shock pulsar wind nebula}.
\newblock {\em arXiv e-prints}, page arXiv:2502.15447, February 2025.

\bibitem{LHAASOJ1849}
Zhen {Cao}, F.~{Aharonian}, Y.~X. {Bai}, et~al.
\newblock {An extreme particle accelerator powered by pulsar PSR J1849-0001}.
\newblock {\em Nature Astronomy}, April 2026.

\bibitem{aharonian2006first}
F~Aharonian, AG~Akhperjanian, AR~Bazer-Bachi, et~al.
\newblock First detection of a vhe gamma-ray spectral maximum from a cosmic
  source: Hess discovery of the vela x nebula.
\newblock {\em Astronomy \& Astrophysics}, 448(2):L43--L47, 2006.

\bibitem{de2012probing}
P~de~Wilt, N~Maxted, and G~Rowell.
\newblock Probing the extent of the non-thermal emission from the vela x region
  at tev energies with hess.
\newblock 2012.

\bibitem{2006A&A460365A}
F.~{Aharonian}, A.~G. {Akhperjanian}, A.~R. {Bazer-Bachi}, et~al.
\newblock {Energy dependent {\ensuremath{\gamma}}-ray morphology in the pulsar
  wind nebula HESS J1825-137}.
\newblock {\em Astronomy \& Astrophysics}, 460(2):365--374, December 2006.

\bibitem{2019A&A621A116H}
H.~{Abdalla}, F.~{Aharonian}, F.~{Ait Benkhali}, et~al.
\newblock {Particle transport within the pulsar wind nebula HESS J1825-137}.
\newblock {\em Astronomy \& Astrophysics}, 621:A116, January 2019.

\bibitem{2012A&A548A46H}
A.~{Abramowski}, F.~{Acero}, F.~{Aharonian}, et~al.
\newblock {Identification of HESS J1303-631 as a pulsar wind nebula through
  {\ensuremath{\gamma}}-ray, X-ray, and radio observations}.
\newblock {\em Astronomy \& Astrophysics}, 548:A46, December 2012.

\bibitem{2017Sci358911A}
A.~U. {Abeysekara}, A.~{Albert}, R.~{Alfaro}, et~al.
\newblock {Extended gamma-ray sources around pulsars constrain the origin of
  the positron flux at Earth}.
\newblock {\em Science}, 358(6365):911--914, November 2017.

\bibitem{2021PhRvL126x1103A}
F.~{Aharonian}, Q.~{An}, L.~X. {Axikegu}, Bai, et~al.
\newblock {Extended Very-High-Energy Gamma-Ray Emission Surrounding PSR J 0622
  +3749 Observed by LHAASO-KM2A}.
\newblock {\em PRL}, 126(24):241103, June 2021.

\bibitem{ray2011precise}
Paul~S Ray, M~Kerr, D~Parent, AA~Abdo, L~Guillemot, SM~Ransom, N~Rea, MT~Wolff,
  A~Makeev, MSE Roberts, et~al.
\newblock Precise $\gamma$-ray timing and radio observations of 17 fermi
  $\gamma$-ray pulsars.
\newblock {\em The Astrophysical Journal Supplement Series}, 194(2):17, 2011.

\bibitem{2009Sci325840A}
A.~A. {Abdo}, M.~{Ackermann}, M.~{Ajello}, et~al.
\newblock {Detection of 16 Gamma-Ray Pulsars Through Blind Frequency Searches
  Using the Fermi LAT}.
\newblock {\em Science}, 325(5942):840, August 2009.

\bibitem{2020ApJS24733A}
S.~{Abdollahi}, F.~{Acero}, M.~{Ackermann}, et~al.
\newblock {Fermi Large Area Telescope Fourth Source Catalog}.
\newblock {\em ApJS}, 247(1):33, March 2020.

\bibitem{barral2018extreme}
Alba~Fern{\'a}ndez Barral.
\newblock {\em Extreme particle acceleration in microquasar jets and pulsar
  wind nebulae with the MAGIC telescopes}.
\newblock Springer, 2018.

\bibitem{amenomori2009observation}
M~Amenomori, XJ~Bi, D~Chen, et~al.
\newblock Observation of tev gamma rays from the fermi bright galactic sources
  with the tibet air shower array.
\newblock {\em The Astrophysical Journal Letters}, 709(1):L6, 2009.

\bibitem{2014APh5716A}
A.~A. {Abdo}, A.~U. {Abeysekara}, B.~T. {Allen}, et~al.
\newblock {Milagro observations of potential TeV emitters}.
\newblock {\em Astroparticle Physics}, 57:16--25, May 2014.

\bibitem{2021Sci373425L}
Zhen {Cao}, F.~{Aharonian}, Q.~{An}, et~al.
\newblock {Peta-electron volt gamma-ray emission from the Crab Nebula}.
\newblock {\em Science}, 373:425--430, July 2021.

\bibitem{he2018design}
Huihai He and LHAASO collaboration.
\newblock Design of the lhaaso detectors.
\newblock {\em Radiation Detection Technology and Methods}, 2:1--8, 2018.

\bibitem{2021ChPhC45b5002A}
F.~{Aharonian}, Q.~{An}, {Axikegu}, et~al.
\newblock {Observation of the Crab Nebula with LHAASO-KM2A - a performance
  study}.
\newblock {\em Chinese Physics C}, 45(2):025002, February 2021.

\bibitem{2021ChPhC45h5002A}
F.~{Aharonian}, Q.~{An}, {Axikegu}, et~al.
\newblock {Performance of LHAASO-WCDA and observation of the Crab Nebula as a
  standard candle}.
\newblock {\em Chinese Physics C}, 45(8):085002, August 2021.

\bibitem{cao2024lhaaso}
Zhen Cao, F~Aharonian, Q~An, et~al.
\newblock Lhaaso-km2a detector simulation using geant4.
\newblock {\em Radiation Detection Technology and Methods}, pages 1--11, 2024.

\bibitem{2023ApJ958191S}
D.~A. {Smith}, S.~{Abdollahi}, M.~{Ajello}, et~al.
\newblock {The Third Fermi Large Area Telescope Catalog of Gamma-Ray Pulsars}.
\newblock {\em ApJ}, 958(2):191, December 2023.

\bibitem{brownsberger2014survey}
Sasha Brownsberger and Roger~W Romani.
\newblock A survey for h$\alpha$ pulsar bow shocks.
\newblock {\em The Astrophysical Journal}, 784(2):154, 2014.

\bibitem{2026Fast}
Jianli {Zhang}, Hui {Zhu}, Guanhong {Lin}, et~al.
\newblock {FAST Discovery of $\mu$Jy Radio Pulsations from PSR J2238+5903,
  Providing a DM Distance Anchor for the Candidate TeV Halo 1LHAASO
  J2238+5900}.
\newblock 2026.

\bibitem{Su_2019}
Yang Su, Ji~Yang, Shaobo Zhang, et~al.
\newblock The milky way imaging scroll painting (mwisp): Project details and
  initial results from the galactic longitudes of 25.°8–49.°7.
\newblock {\em The Astrophysical Journal Supplement Series}, 240(1):9, jan
  2019.

\bibitem{2017ApJ83457M}
Marc-Antoine {Miville-Desch{\^e}nes}, Norman {Murray}, and Eve~J. {Lee}.
\newblock {Physical Properties of Molecular Clouds for the Entire Milky Way
  Disk}.
\newblock {\em ApJ}, 834(1):57, January 2017.

\bibitem{2015ICRC34922Z}
V.~{Zabalza}.
\newblock {Naima: a Python package for inference of particle distribution
  properties from nonthermal spectra}.
\newblock In {\em 34th International Cosmic Ray Conference (ICRC2015)},
  volume~34 of {\em International Cosmic Ray Conference}, page 922, July 2015.

\bibitem{2014ApJ783100K}
D.~{Khangulyan}, F.~A. {Aharonian}, and S.~R. {Kelner}.
\newblock {Simple Analytical Approximations for Treatment of Inverse Compton
  Scattering of Relativistic Electrons in the Blackbody Radiation Field}.
\newblock {\em ApJ}, 783(2):100, March 2014.

\bibitem{2010PhRvD82d3002A}
F.~A. {Aharonian}, S.~R. {Kelner}, and A.~Yu. {Prosekin}.
\newblock {Angular, spectral, and time distributions of highest energy protons
  and associated secondary gamma rays and neutrinos propagating through
  extragalactic magnetic and radiation fields}.
\newblock {\em PRD}, 82(4):043002, August 2010.

\bibitem{2022MNRAS5092339N}
Giovanni {Natale}, Cristina~C. {Popescu}, Mark {Rushton}, et~al.
\newblock {A radiation transfer model for the Milky Way: II. The global
  properties and large-scale structure}.
\newblock {\em MNRAS}, 509(2):2339--2361, January 2022.

\bibitem{2009ApJ69412M}
F.~{Mattana}, M.~{Falanga}, D.~{G{\"o}tz}, et~al.
\newblock {The Evolution of the {\ensuremath{\gamma}}- and X-Ray Luminosities
  of Pulsar Wind Nebulae}.
\newblock {\em ApJ}, 694(1):12--17, March 2009.

\bibitem{2019PhRvL123v1103L}
Ruo-Yu {Liu}, Huirong {Yan}, and Heshou {Zhang}.
\newblock {Understanding the Multiwavelength Observation of Geminga's Tev Halo:
  The Role of Anisotropic Diffusion of Particles}.
\newblock {\em PRL}, 123(22):221103, November 2019.

\bibitem{2024ApJ9699W}
Qi-Zuo {Wu}, Chao-Ming {Li}, Xuan-Han {Liang}, Chong {Ge}, and Ruo-Yu {Liu}.
\newblock {Diagnosing the Particle Transport Mechanism in the Pulsar Halo via
  X-Ray Observations}.
\newblock {\em ApJ}, 969(1):9, July 2024.

\bibitem{2022FrASS922100F}
Kun {Fang}.
\newblock {Gamma-ray pulsar halos in the Galaxy}.
\newblock {\em Frontiers in Astronomy and Space Sciences}, 9:1022100, October
  2022.

\bibitem{2011ApJ740L26C}
Roger~A. {Chevalier} and Stephen~P. {Reynolds}.
\newblock {Pulsar Wind Nebulae with Thick Toroidal Structure}.
\newblock {\em ApJL}, 740(1):L26, October 2011.

\bibitem{2019ApJ87754B}
Yiwei {Bao}, Siming {Liu}, and Yang {Chen}.
\newblock {On the Gamma-Ray Nebula of Vela Pulsar. I. Very Slow Diffusion of
  Energetic Electrons within the TeV Nebula}.
\newblock {\em ApJ}, 877(1):54, May 2019.

\bibitem{2019ApJ881148B}
Yiwei {Bao} and Yang {Chen}.
\newblock {On the Gamma-Ray Nebula of Vela Pulsar. II. The Soft Spectrum of the
  Extended Radio Nebula}.
\newblock {\em ApJ}, 881(2):148, August 2019.

\bibitem{2025ApJ98098S}
Jiaxu {Sun}, Yang {Chen}, Yiwei {Bao}, Xiao {Zhang}, and Xin {Zhou}.
\newblock {Modeling the Saddle-like GeV-TeV Spectrum of HESS J1809-193: Gamma
  Rays Arising from Reverse-shocked Pulsar Wind Nebula?}
\newblock {\em ApJ}, 980(1):98, February 2025.

\bibitem{2017PhRvD96j3016L}
Tim {Linden}, Katie {Auchettl}, Joseph {Bramante}, et~al.
\newblock {Using HAWC to discover invisible pulsars}.
\newblock {\em PRD}, 96(10):103016, November 2017.

\bibitem{2023ApJ944L29A}
A.~{Albert}, R.~{Alfaro}, J.~C. {Arteaga-Vel{\'a}zquez}, et~al.
\newblock {HAWC Detection of a TeV Halo Candidate Surrounding a Radio-quiet
  Pulsar}.
\newblock {\em ApJL}, 944(2):L29, February 2023.

\bibitem{2024ApJ968117Z}
Dong {Zheng} and Zhongxiang {Wang}.
\newblock {Finding Candidate TeV Halos among Very-high-energy Sources}.
\newblock {\em ApJ}, 968(2):117, June 2024.

\end{thebibliography}

\clearpage
Zhen Cao$^{1,2,3}$,
F. Aharonian$^{3,4,5,6}$,
Y.X. Bai$^{1,3}$,
Y.W. Bao$^{7}$,
D. Bastieri$^{8}$,
X.J. Bi$^{1,2,3}$,
Y.J. Bi$^{1,3}$,
W. Bian$^{7}$,
J. Blunier$^{9}$,
A.V. Butkevich$^{10}$
% ...

(The LHAASO Collaboration)

$^{1}$ State Key Laboratory of Particle Astrophysics \& Experimental Physics Division \& Computing Center, Institute of High Energy Physics, Chinese Academy of Sciences, 100049 Beijing, China\\
$^{2}$ University of Chinese Academy of Sciences, 100049 Beijing, China\\
$^{3}$ Tianfu Cosmic Ray Research Center, 610000 Chengdu, Sichuan,  China\\
$^{4}$ University of Science and Technology of China, 230026 Hefei, Anhui, China\\
$^{5}$ Yerevan State University, 1 Alek Manukyan Street, Yerevan 0025, Armenia \\
$^{6}$ Max-Planck-Institut for Nuclear Physics, P.O. Box 103980, 69029  Heidelberg, Germany\\
$^{7}$ Tsung-Dao Lee Institute \& School of Physics and Astronomy, Shanghai Jiao Tong University, 200240 Shanghai, China\\
$^{8}$ Center for Astrophysics, Guangzhou University, 510006 Guangzhou, Guangdong, China\\
$^{9}$ APC, Universit'e Paris Cit'e, CNRS/IN2P3, CEA/IRFU, Observatoire de Paris, 119 75205 Paris, France\\
$^{10}$ Institute for Nuclear Research of Russian Academy of Sciences, 117312 Moscow, Russia\\
% ...

\end{document}